# Interfacial adhesion between graphene and silicon dioxide by density functional theory with van der Waals corrections

Wei Gao[1], Penghao Xiao[2], Graeme Henkelman[2], Kenneth M. Liechti[1] and Rui Huang[1]

[1]*Department of Aerospace Engineering and Engineering Mechanics, University of Texas at Austin, Austin, TX 78712, USA*

[2]*Department of Chemistry and the Institute for Computational and Engineering Sciences, University of Texas at Austin, Austin, TX 78712, USA*

(Date: March 13, 2014)

## Abstract

Interfacial adhesion between graphene and a $SiO_2$ substrate is studied by density functional theory (DFT) with dispersion corrections. The results demonstrate the van der Waals (vdW) interaction as the predominate mechanism for the graphene/$SiO_2$ interface. It is found that the interaction strength is strongly influenced by changes of the $SiO_2$ surface structures due to surface reactions with water. The adhesion energy is reduced when the reconstructed $SiO_2$ surface is hydroxylated, and further reduced when covered by a monolayer of adsorbed water molecules. Thus, the effect of humidity may help explain the wide variation of adhesion energies measured in recent experiments between graphene and $SiO_2$. Moreover, it is noted that vdW forces are required to accurately model the graphene/$SiO_2$ interface with DFT and that the adhesion energy is underestimated by empirical force fields commonly used in atomistic simulations.

Graphene, a two-dimensional crystal membrane, has drawn tremendous interest due to its remarkable electronic and mechanical properties. With respect to applications such as graphene-based nanoelectronic devices,[1] the interfacial properties between graphene and the supporting substrate are of great importance. Interfacial adhesion energies have been measured for graphene on various substrate materials such as silicon dioxide ($SiO_2$)[2-4] and copper.[5,6] The $SiO_2$ substrate was instrumental for the first experimental observation of mechanically exfoliated graphene[7] and has been widely used as a dielectric medium in integrated circuits. Using a combined SEM/AFM/STM technique, Ishigami et al.[8] showed that monolayer graphene largely follows the underlying morphology of $SiO_2$, and they estimated the adhesion energy between graphene and $SiO_2$ to be 0.096 $J/m^2$, based on the interlayer van der Waals (vdW) interaction in bulk graphite. However, the measurements by Koenig et al.[2] reported a strong adhesion of 0.45 $J/m^2$ between graphene and the $SiO_2$ substrate. More recently, a similar experiment yielded a considerably lower adhesion energy of 0.24 $J/m^2$.[3] It was suggested that the difference could arise from the surface properties of $SiO_2$, such as surface roughness and chemical reactivity. The effect of surface roughness, which has been analyzed using a macroscopic continuum model,[9-11] may contribute to the experimental variations. In this paper, the influence of the microscopic surface structures and their chemical reactivity on interfacial adhesion is investigated using density functional theory (DFT).

DFT calculations of graphene on $SiO_2$ have been reported previously. While $SiO_2$ is typically amorphous in experiments, DFT calculations are generally limited to crystalline $SiO_2$, with only a few exceptions.[12,13] Among the crystalline $SiO_2$ phases, α-quartz is the most stable under ambient conditions. Several DFT studies reported that C-O and C-Si covalent bonds can form at the graphene/$SiO_2$ interface due to the reactivity of dangling bonds on the $SiO_2$ surface.[14-16] As a result, a strong interfacial adhesion between graphene and $SiO_2$ was predicted. For instance, Hossain[16] calculated the adhesion energy as 62.10 meV/Å$^2$ (or equivalently, 0.995 J/m$^2$) for the O-terminated $SiO_2$ surface. However, many experiments[2,3,8,17] suggested that the interaction between graphene and $SiO_2$ is physisorption in nature, dominated by vdW interactions rather than covalent bonds. In fact, previous studies[18,19] on α-quartz have shown that the cleaved (001) surface undergoes a reconstruction at around 300 K to become O-terminated with six-membered rings as shown in Fig. 1(a). Meanwhile, the under-coordinated (001) surface is hydrophilic, which commonly reacts with ambient water to yield silanol groups (Si-OH). The hydroxylated α-quartz surface is characterized by a zigzag network with alternating strong and weak hydrogen bonds as shown in Fig. 1(b). Cuong et al.[20] studied both the reconstructed and hydroxylated α-quartz surfaces using DFT with the local density approximation (LDA). They obtained a binding energy of 14.6 meV per C atom (equivalent to an adhesion energy of 0.090 J/m$^2$) for the reconstructed surface and 12.8 meV per C atom (0.079 J/m$^2$) for the hydroxylated surface. Noticing that LDA does not take into account the dispersive interactions, Fan et al.[21] considered vdW interactions with a semiempirical approach (DFT-D2) and obtained an adhesion energy of 0.235 J/m$^2$ for the reconstructed α-quartz surface. Recently, several other methods have been proposed to account for vdW interactions in DFT calculations including approaches by Tkatchenko and Scheffler[22,23] (vdW-TS) and Klimes et al.[24] (optPBE-vdW). In this paper, we compare different DFT methods for interfacial adhesion between graphene and $SiO_2$ with different surface microstructures. In addition to the reconstructed and hydroxylated surfaces, water absorption on the surface is also considered, since the silano groups on the hydroxylated surface are sensitive to the adsorption of small molecules such as $H_2O$ under ambient conditions. In particular, the adsorption of water on the α-quartz surface was found to be thermodynamically favorable in previous studies.[25-27] DFT calculations have shown that, when water is adsorbed on the hydroxylated surface, the weak hydrogen bonds are broken and new hydrogen bonds are formed between the hydroxyl groups and water molecules.[27,28] When the coverage of water molecules reaches one monolayer, a hexagonal $H_2O$ network similar to the basal plane of ice Ih is formed on the surface, as shown in Fig. 1(c).

All the DFT calculations in this study were performed using the plane-wave-based Vienna ab initio simulation package (VASP).[28,29] Projector augmented wave (PAW)[30,31] pseudopotentials were used to represent ionic cores, and the electronic kinetic energy cutoff for the plane-wave basis describing the valence electrons was set to 520 eV. A $4 \times 4 \times 1$ $k$–point mesh was used for structure relaxation and a $14 \times 14 \times 1$ $k$–point mesh for self-consistent static calculation. The ground state structural parameters of bulk $SiO_2$ and graphene were first calculated using the five DFT methods listed in Table I. It is found that the calculated structure is over-bound with LDA and slightly under-bound by the other methods, as compared to experiments.[32,33] The supercell for the adhesion energy calculations consisted of a $2 \times 2$ graphene sheet on a $1 \times 1$ $SiO_2$ unit cell with a vacuum layer of 20 Å thickness separating the periodic images of the slab. The in-plane dimension of the supercell was set by the equilibrium lattice constant of graphene. The lattice constant of the $SiO_2$ substrate was adjusted by a biaxial strain to accommodate the lattice mismatch, as listed in Table I. To compute the adhesion energy, the system was fully relaxed, except for the middle layer in the $SiO_2$ slab, which was frozen in the bulk structure. The adhesion energy $E_{ad}$ was then calculated by

$$E_{ad} = E_g + E_s - E_{g/s}, \qquad (1)$$

where $E_g$, $E_s$ and $E_{g/s}$ are energies of isolated graphene, isolated $SiO_2$ substrate, and the graphene/$SiO_2$ system, respectively. It is noted that different binding positions could be obtained by shifting the relative locations between graphene and SiO$_2$ along the lattice vector directions, with a periodicity same as the primitive cell of graphene. As shown in Fig. 2(a), we partition the primitive cell of graphene into a 6×6 equal spaced mesh, so that the adhesion energy could be calculated at 36 different relative positions. The most stable configuration corresponds to the one with the lowest energy $E_{g/s}$, with which the adhesion energy is calculated.

Table II lists the adhesion energies from our calculations. The generalized gradient approximation with Perdew-Burke-Ernzerhof functional (GGA-PBE)[34] yields minimal adhesion for all the surface types considered, which is expected, since no vdW interactions are accounted for. It has been shown that the local density approximation (LDA)[35] is able to predict correct interlayer distances for some layered materials including graphite. However, it is purely local and hence not able to fully describe long-range dispersion interactions. Our results indicate that LDA considerably underestimates the adhesion energy in comparison with the experimental measurements, although the predicted equilibrium separation is very close. Previous studies have shown the importance of vdW corrections to traditional DFT for describing the interfaces in graphene-based systems such as graphite,[36,37] graphene on metals substrates[38] and graphene on SiC.[39] Many schemes have been proposed for correcting DFT calculations with dispersion effects for vdW interactions, among which the DFT-D2, vdW-TS, and optPBE-vdW methods are used in the present study. The DFT-D2 method[40] adds a pairwise interatomic $C_{6AB}R_{AB}^{-6}$ term to the conventional Kohn-Sham energy, where $R_{AB}$ is the distance between atoms $A$ and $B$, and $C_{6AB}$ is the corresponding coefficient. As shown in our results, this correction brings in appreciable adhesion energy between graphene and SiO$_2$. The drawback of DFT-D2 is its empirical nature, since the pairwise coefficients in the correction term are obtained by fitting to experiments or post-Hartree-Fock analysis, with the requirement of being independent of the chemical environment. Tkatchenko and Scheffler proposed a more sophisticated method (vdW-TS) to compute the $C_{6AB}$ coefficients from the mean-field ground-state electron density of molecules and solids.[22,23] Our calculations show that the adhesion energies from vdW-TS are about 50% greater than those from DFT-D2. Another vdW corrected DFT method is optPBE-vdW,[24] which uses the nonlocal correlation description from the nonempirical and electron density based Chalmers-Rutgers vdW-DF[41] method but with its exchange functional optimized based on S22 datasets.[42] It is found that the adhesion energy predicted by optPBE-vdW compares closely to the prediction by vdW-TS, and both are in good agreement with experimental measurements.[2,3]

Fig. 2 (b)-(d) illustrates the optimized binding structures of graphene on the three types of $SiO_2$ surfaces. It is noted that the most stable binding structure does not depend on the choice of DFT method. Moreover, the energy variations among the 36 binding locations are 6-10% of the total adhesion energies, indicating that the binding between graphene and the $SiO_2$ substrate is insensitive to their relative positions. It is found that the adhesion energy is reduced by surface hydroxylation and further reduced by adsorption of a water monolayer. In the latter case, the adhesion energy includes contributions from both graphene-water and graphene-$SiO_2$ interactions. The graphene-water interaction was investigated previously by first-principle calculations, which calculated the adsorption energy between a water monomer and graphene to be 90 meV/$H_2O$.[43] With the number density of water molecules on the $SiO_2$ surface in our calculation (~9.33 nm$^{-2}$), the graphene-water interaction would contribute about 0.134 J/m$^2$ toward the total adhesion energy of 0.210 J/m$^2$. The contribution from the graphene-$SiO_2$ interaction is then 0.076 J/m$^2$, which is much lower than the adhesion energy between graphene and a bare $SiO_2$. The presence of the water monolayer thus weakens the vdW interaction between graphene and $SiO_2$, which may be partly attributed to the relatively large separation between graphene and $SiO_2$ (~5.06 Å). While the full hydroxylation and monolayer water coverage of the surface are considered here, the density of silanol groups or water adsorption for a real $SiO_2$ surface would depend on the ambient conditions, such as the relative humidity. Nevertheless, our study suggests that microstructural changes due to the chemical reactivity of the $SiO_2$ surface with water may contribute to the variation of adhesion energies measured in experiment,[2,3] in addition to the macroscopic effects due to surface roughness. We note that the macroscopic capillary effect is not considered in this study, which may become important at relatively high humidity and give rise to different characteristics of adhesion.[25]

In all cases, it is found that graphene maintains its planar configuration on top of the $SiO_2$ substrate. This is expected for two reasons: the substrate surface is atomically smooth (i.e., no macroscopic roughness is considered) and no temperature effect is taken into account in the DFT calculations (hence no thermal rippling). As a result, the separation ($\delta$) between graphene and the substrate can be defined as the distance between the C atoms in graphene and the topmost atoms of the substrate (including water molecules) as shown in Fig. 3(a). By freezing the out-of-plane displacement of graphene, the interaction energy $U$ can be calculated at different separations; the minimum interaction energy is reached at the equilibrium separation ($\delta_0$). The function $U(\delta)$, calculated using the vdW-TS method, is plotted in Fig. 3(a) for three different surface structures. In all three cases, the interaction energy functions show long-range tails, revealing the nature of dispersion interactions. As a simple mathematical model, the Lennard-Jones (LJ) potential is commonly used in atomistic simulations based on empirical force fields to account for the dispersion forces, including the graphene/SiO2 interface.[44,45] The LJ potential between atoms $i$ and $j$ can be written as

$$V_{ij}(R_{ij}) = \varepsilon_{ij}\left[\left(\frac{\sigma_{ij}}{R_{ij}}\right)^{12} - \left(\frac{\sigma_{ij}}{R_{ij}}\right)^{6}\right], \tag{2}$$

where $R_{ij}$ is the atomic distance, $\sigma_{ij}$ and $\varepsilon_{ij}$ are the pairwise parameters. By integrating Eq. (2) with respect to all atoms, the interaction energy between graphene and $SiO_2$ substrate per unit area can be obtained as[9]

$$U_{LJ}(\delta) = \sum_{j} \frac{2\pi\rho_{j}\varepsilon_{ij}}{A_0}\left(\frac{\sigma_{ij}^{12}}{45\delta^{9}} - \frac{\sigma_{ij}^{6}}{6\delta^{3}}\right), \tag{3}$$

where the subscript $i$ represents a C atom, $j$ represents Si or O, $\rho_j$ is the number density of Si or O atoms in the substrate ($\rho_{Si} = 25.0\text{nm}^{-3}$ and $\rho_{O} = 50.0\text{nm}^{-3}$), and $A_0$ is the area of a unit cell of graphene. The summation in Eq. (3) takes both Si-C and O-C interactions into account. In the empirical force field, the parameters $\sigma_{ij}$ and $\varepsilon_{ij}$ for each pairwise interaction are obtained by fitting to experiments or first principle calculations. Considering three typical force fields (UFF,[46] Charmm[47] and Dreiding[48]), we calculated the interaction energy $U_{LJ}(\delta)$ using Eq. (3) for the reconstructed $SiO_2$ surface, as shown in Fig. 3(b). Apparently, the equilibrium separation between graphene and $SiO_2$ is close to the DFT result, but the adhesion energy is underestimated by all of the empirical methods. For the hydroxylated and water monolayer covered $SiO_2$ surfaces, the use of empirical force fields would be more problematic.

To further investigate the interfacial interaction and its potential impact on the physical properties of graphene, we calculated the electronic structures of the graphene/$SiO_2$ system. The band structures obtained from the vdW-TS method are shown in Fig. 4. The shape of the Dirac cone of the pristine monolayer graphene is preserved for all three surfaces with tiny band gaps at the K point. The band gap opening can be understood by the breaking of the sublattice symmetry of graphene due to its interaction with the substrate. Such a mechanism has a more significant effect on band gap opening of graphene on SiC[49] and hexagonal boron nitride[50] substrates, but the effect is negligible for the graphene/SiO$_2$ system since the band gap is much less than the thermal energy at room temperature (~25 meV). Moreover, it is noted that there is no Fermi level shift in the three systems, indicating no significant charge transfer induced electrostatic interactions. Previous experiments[51,52] have shown some evidence for both p-type and n-type doping of graphene on SiO$_2$ substrates, which may be accounted for by including non-ideal aspects of the system such as surface defects and other environmental effects. Based on an analytical model, Sabio et al.[53] studied electrostatic interactions between graphene and SiO$_2$ along with other materials (including water molecules) in its environment. They found that the leading electrostatic interactions arise from the surface polar modes of SiO$_2$ and electrical dipoles of water molecules, with estimated interaction energies of 0.4 meV/Å$^2$ (0.0064 J/m$^2$) and 1 meV/Å$^2$ (0.016 J/m$^2$), respectively; both are significantly lower than the adhesion energies due to the vdW interactions in the present DFT calculations.

In conclusion, the interfacial adhesion between graphene and SiO$_2$ substrate is studied by DFT methods with vdW interactions. It is found that the interaction between graphene and SiO$_2$ is dominated by dispersion forces. The adhesion energy is reduced by surface hydroxylation and further reduced by adsorption of water molecules. Our study shows that the vdW-TS and optPBE-vdW methods are both suitable for studying the interactions between graphene and SiO$_2$. Moreover, the discrepancy between DFT and empirical force fields suggests a need for more accurate parameters to describe the graphene/SiO$_2$ system. Finally, it is found that the vdW interactions have negligible influence on the electronic band structure of graphene.

**Acknowledgments**

The authors gratefully acknowledge financial support of this work by the National Science Foundation through Grant No. CMMI-1130261. The authors acknowledge the Texas Advanced Computing Center (TACC) at the University of Texas at Austin for providing HPC resources that have contributed to the research results reported within this paper.

**References**

[1] Y. Q. Wu, D. B. Farmer, F. N. Xia, and P. Avouris, Proc. IEEE **101**, 1620 (2013).
[2] S. P. Koenig, N. G. Boddeti, M. L. Dunn, and J. S. Bunch, Nature Nanotech. **6**, 543 (2011).
[3] N. G. Boddeti, S. P. Koenig, R. Long, J. L. Xiao, J. S. Bunch, and M. L. Dunn, J. Appl. Mech. **80**, 040909 (2013).
[4] Z. Zong, C. L. Chen, M. R. Dokmeci, and K. T. Wan, J. Appl. Phys. **107**, 026104 (2010).
[5] T. Yoon, W. C. Shin, T. Y. Kim, J. H. Mun, T. S. Kim, and B. J. Cho, Nano Lett. **12**, 1448 (2012).
[6] Z. Cao, P. Wang, W. Gao, L. Tao, J. W. Suk, R. S. Ruoff, D. Akinwande, R. Huang, and K. M. Liechti, Carbon **69**, 390 (2014).
[7] K. S. Novoselov, A. K. Geim, S. V. Morozov, D. Jiang, Y. Zhang, S. V. Dubonos, I. V. Grigorieva, and A. A. Firsov, Science **306**, 666 (2004).
[8] M. Ishigami, J. H. Chen, W. G. Cullen, M. S. Fuhrer, and E. D. Williams, Nano Lett. **7**, 1643 (2007).
[9] Z. H. Aitken and R. Huang, J. Appl. Phys. **107**, 123531 (2010).
[10] W. Gao and R. Huang, J. Phys. D: Appl. Phys. **44**, 452001 (2011).
[11] T. Li and Z. Zhang, J. Phys. D: Appl. Phys. **43**, 075303 (2010).
[12] R. H. Miwa, T. M. Schmidt, W. L. Scopel, and A. Fazzio, Appl. Phys. Lett. **99**, 163108 (2011).
[13] A. N. Rudenko, F. J. Keil, M. I. Katsnelson, and A. I. Lichtenstein, Phys. Rev. B **84**, 085438 (2011).
[14] Y. J. Kang, J. Kang, and K. J. Chang, Phys. Rev. B **78**, 115404 (2008).
[15] P. Shemella and S. K. Nayak, Appl. Phys. Lett. **94**, 032101 (2009).
[16] M. Z. Hossain, Appl. Phys. Lett. **95**, 143125 (2009).
[17] E. Stolyarova, K. T. Rim, S. M. Ryu, J. Maultzsch, P. Kim, L. E. Brus, T. F. Heinz, M. S. Hybertsen, and G. W. Flynn, Proc. Natl. Acad. Sci. U.S.A. **104**, 9209 (2007).
[18] G. M. Rignanese, A. De Vita, J. C. Charlier, X. Gonze, and R. Car, Phys. Rev. B **61**, 13250 (2000).
[19] T. P. M. Goumans, A. Wander, W. A. Brown, and C. R. A. Catlow, Phys. Chem. Chem. Phys. **9**, 2146 (2007).
[20] N. T. Cuong, M. Otani, and S. Okada, Phys. Rev. Lett. **106**, 106801 (2011).
[21] X. F. Fan, W. T. Zheng, V. Chihaia, Z. X. Shen, and J. L. Kuo, J. Phys. : Condens. Matter **24**, 305004 (2012).
[22] T. Bucko, S. Lebegue, J. Hafner, and J. G. Angyan, Phys. Rev. B **87**, 064110 (2013).
[23] A. Tkatchenko and M. Scheffler, Phys. Rev. Lett. **102**, 073005 (2009).
[24] J. Klimes, D. R. Bowler, and A. Michaelides, J. Phys. : Condens. Matter **22**, 022201 (2010).
[25] Arthur W Adamson and Alice Petry Gast, Physical chemistry of surfaces. (Wiley New York, 1990).
[26] Y. W. Chen and H. P. Cheng, Appl. Phys. Lett. **97**, 161909 (2010).
[27] J. J. Yang and E. G. Wang, Phys. Rev. B **73**, 035406 (2006).
[28] G. Kresse and J. Furthmuller, Phys. Rev. B **54**, 11169 (1996).
[29] G. Kresse and J. Hafner, Phys. Rev. B **47**, 558 (1993).
[30] G. Kresse and D. Joubert, Phys. Rev. B **59**, 1758 (1999).
[31] P. E. Blochl, Phys. Rev. B **50**, 17953 (1994).
[32] Y. Baskin and L. Meyer, Phys. Rev. **100**, 544 (1955).
[33] G. Will, M. Bellotto, W. Parrish, and M. Hart, J. Appl. Crystallogr. **21**, 182 (1988).
[34] J. P. Perdew, K. Burke, and M. Ernzerhof, Phys. Rev. Lett. **77**, 3865 (1996).
[35] J. P. Perdew and A. Zunger, Phys. Rev. B **23**, 5048 (1981).
[36] H. Rydberg, M. Dion, N. Jacobson, E. Schroder, P. Hyldgaard, S. I. Simak, D. C. Langreth, and B. I. Lundqvist, Phys. Rev. Lett. **91**, 126402 (2003).
[37] B. I. Lundqvist, A. Bogicevic, K. Carling, S. V. Dudiy, S. Gao, J. Hartford, P. Hyldgaard, N. Jacobson, D. C. Langreth, N. Lorente, S. Ovesson, B. Razaznejad, C. Ruberto, H. Rydberg, E. Schroder, S. I. Simak, G. Wahnstrom, and Y. Yourdshahyan, Surf Sci. **493**, 253 (2001).

[38] M. Vanin, J. J. Mortensen, A. K. Kelkkanen, J. M. Garcia-Lastra, K. S. Thygesen, and K. W. Jacobsen, Phys. Rev. B **81**, 081408 (2010).

[39] L. Nemec, V. Blum, P. Rinke, and M. Scheffler, Phys. Rev. Lett. **111**, 065502 (2013).

[40] S. Grimme, J. Comput. Chem. **27**, 1787 (2006).

[41] M. Dion, H. Rydberg, E. Schroder, D. C. Langreth, and B. I. Lundqvist, Phys. Rev. Lett. **92**, 246401 (2004).

[42] P. Jurecka, J. Sponer, J. Cerny, and P. Hobza, Phys. Chem. Chem. Phys. **8**, 1985 (2006).

[43] J. Ma, A. Michaelides, D. Alfe, L. Schimka, G. Kresse, and E. G. Wang, Phys. Rev. B **84**, 033402 (2011).

[44] E. P. Bellido and J. M. Seminario, J Phys Chem C **114**, 22472 (2010).

[45] E. Paek and G. S. Hwang, J. Appl. Phys. **113**, 164901 (2013).

[46] A. K. Rappe, C. J. Casewit, K. S. Colwell, W. A. Goddard, and W. M. Skiff, J. Am. Chem. Soc. **114**, 10024 (1992).

[47] E. R. Cruz-Chu, A. Aksimentiev, and K. Schulten, J. Phys. Chem. B **110**, 21497 (2006).

[48] S. L. Mayo, B. D. Olafson, and W. A. Goddard, J. Phys. Chem. **94**, 8897 (1990).

[49] S. Y. Zhou, G. H. Gweon, A. V. Fedorov, P. N. First, W. A. De Heer, D. H. Lee, F. Guinea, A. H. C. Neto, and A. Lanzara, Nat Mater **6**, 770 (2007).

[50] G. Giovannetti, P. A. Khomyakov, G. Brocks, P. J. Kelly, and J. van den Brink, Phys. Rev. B **76**, 073103 (2007).

[51] H. E. Romero, N. Shen, P. Joshi, H. R. Gutierrez, S. A. Tadigadapa, J. O. Sofo, and P. C. Eklund, Acs Nano **2**, 2037 (2008).

[52] S. Ryu, L. Liu, S. Berciaud, Y. J. Yu, H. T. Liu, P. Kim, G. W. Flynn, and L. E. Brus, Nano Lett. **10**, 4944 (2010).

[53] J. Sabio, C. Seoanez, S. Fratini, F. Guinea, A. H. Castro, and F. Sols, Phys. Rev. B **77**, 195409 (2008).

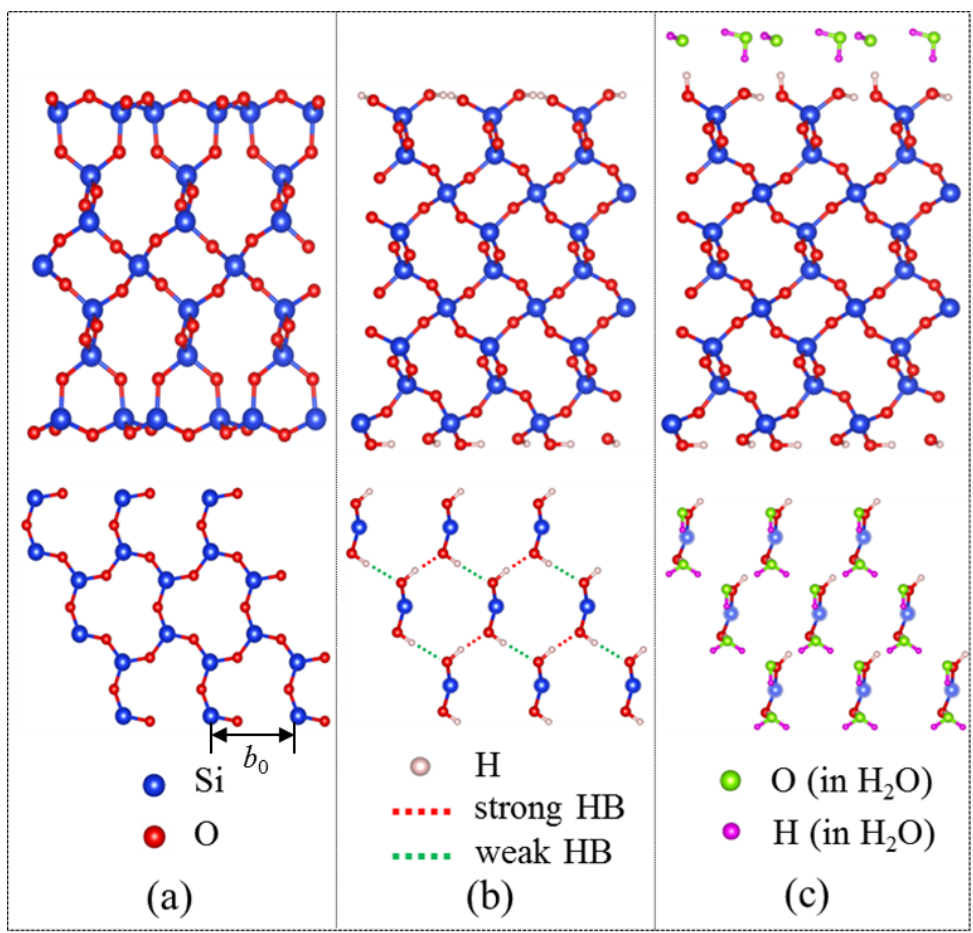

**Figure 1**. Side and top views of the SiO$_2$ substrate with different surface structures: (a) reconstructed, (b) hydroxylated, and (c) covered by a monolayer of water molecules.

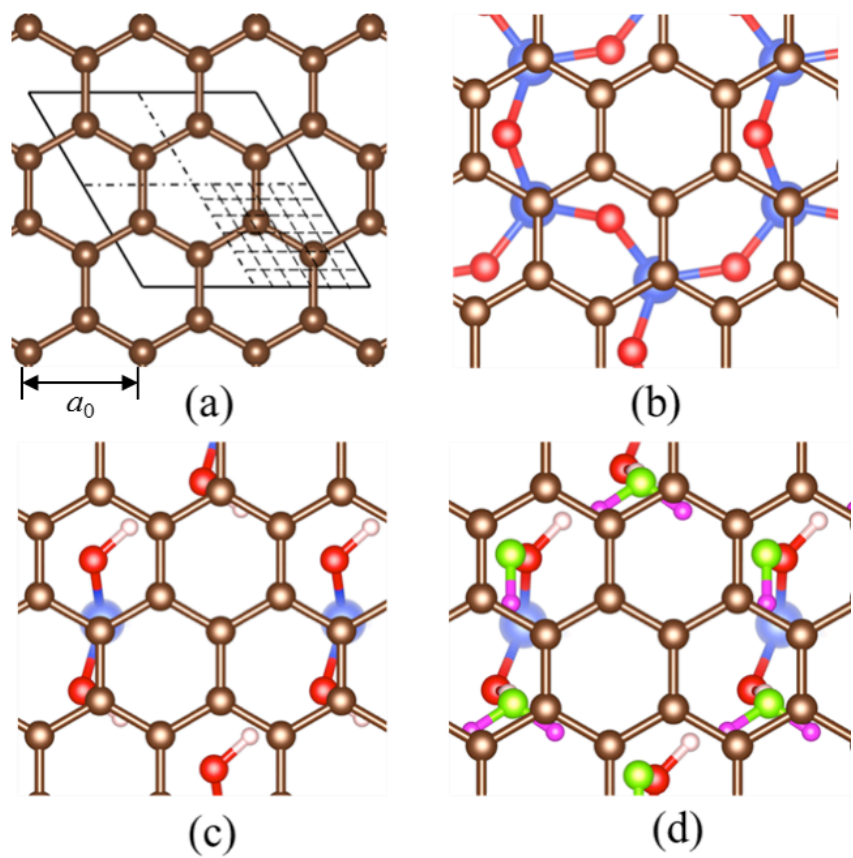

**Figure 2**. (a) Graphene lattice, with a 2x2 unit cell indicated by the parallelogram box, within which a primitive unit cell is partitioned into a 6x6 mesh. (b-d) Top views of the equilibrium structures for graphene on $SiO_2$ with different surface structures: (b) reconstructed, (c) hydroxylated, and (d) covered by a monolayer of water molecules.

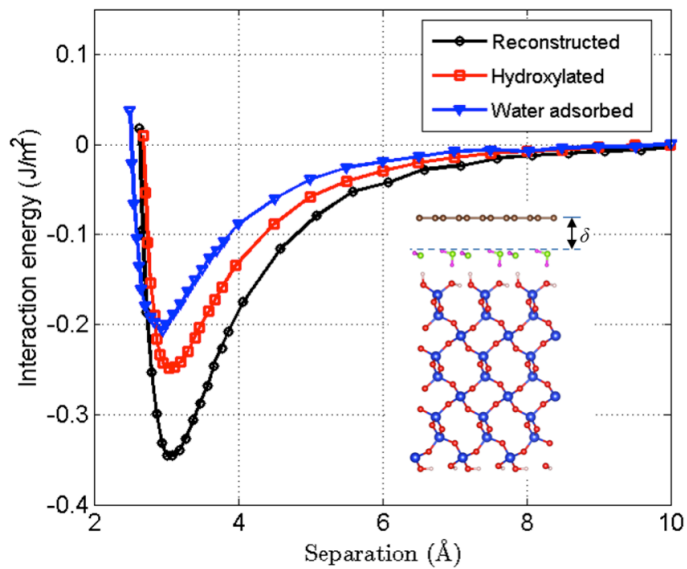

(a)

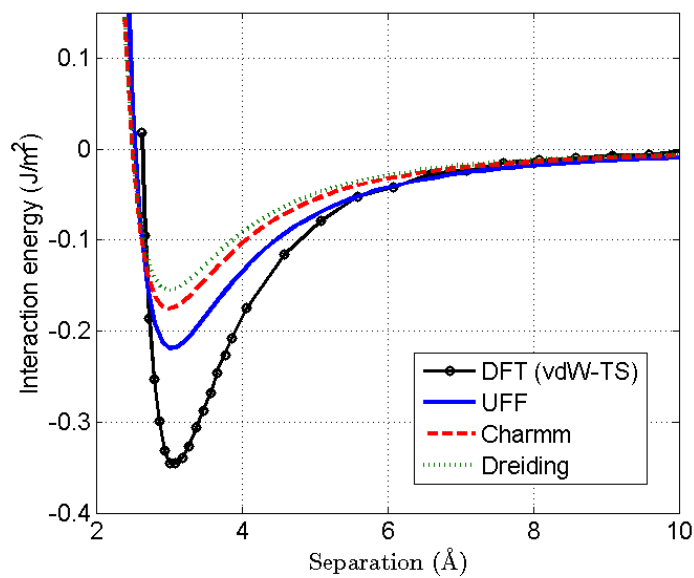

(b)

**Figure 3**. (a) Interaction energy as a function of separation between graphene and $SiO_2$, calculated with the vdW-TS method for three surface structures. The inset shows the side view of graphene on $SiO_2$ with a water monolayer. (b) Comparison of the interaction energy calculated from vdW-TS and three empirical force fields for graphene on $SiO_2$ with reconstructed surface.

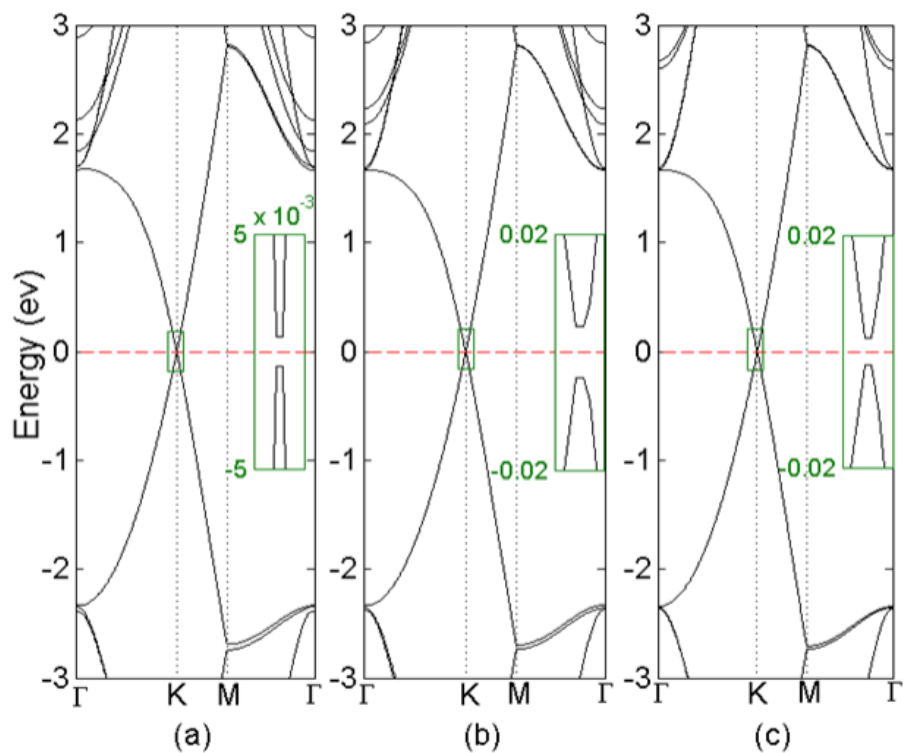

**Figure 4**. Electronic band structures of graphene on $SiO_2$ with different surface structures: (a) reconstructed, (b) hydroxylated, and (c) covered by a monolayer of water molecules. The insets show the band gap around the K point.

Table I. Comparison of lattice parameters for graphene and SiO$_2$ obtained from different DFT methods. The lattice mismatch between graphene and SiO$_2$ is calculated as $\varepsilon_0 = (2a_0 - b_0)/b_0$.

| Method | Graphene, $a_0$ (Å) | SiO$_2$, $b_0$ (Å) | Lattice mismatch, $\varepsilon_0$ (%) |
|---|---|---|---|
| LDA | 2.4462 | 4.8906 | 0.037 |
| GGA-PBE | 2.4678 | 5.0371 | -2.02 |
| DFT-D2 | 2.4685 | 4.9259 | 0.23 |
| vdW-TS | 2.4656 | 4.9764 | -0.91 |
| optPBE-vdW | 2.4713 | 4.9891 | -0.93 |
| Experimental | 2.4589[32] | 4.9124[33] | 0.11 |

Table II. Adhesion energy ($E_{ad}$) and equilibrium separation ($\delta_0$) between graphene and $SiO_2$ with reconstructed, hydroxylated, and water monolayer covered surfaces, obtained from different DFT methods.

| Method | Reconstructed $E_{ad}$ (J/m$^2$) / $\delta_0$ (Å) | Hydroxylated $E_{ad}$ (J/m$^2$) / $\delta_0$ (Å) | Water adsorption $E_{ad}$ (J/m$^2$) / $\delta_0$ (Å) |
| --- | --- | --- | --- |
| GGA-PBE | 0.0027 / 3.556 | 0.0055 / 3.420 | 0.0034 / 3.207 |
| LDA | 0.115 / 3.000 | 0.094 / 3.002 | 0.096 / 2.882 |
| DFT-D2 | 0.229 / 3.006 | 0.166 / 3.043 | 0.134 / 2.800 |
| vdW-TS | 0.349 / 3.089 | 0.242 / 3.164 | 0.210 / 2.993 |
| optPBE-vdW | 0.311 / 3.069 | 0.258 / 3.036 | 0.224 / 2.883 |